\documentclass{article}
\usepackage[sorting=none]{biblatex}
\usepackage{amsmath,amssymb,amsfonts}
\usepackage{algorithmic}
\usepackage{listings}
\usepackage{textcomp}
\usepackage{xcolor}
\usepackage{color}
\usepackage{graphicx}
\usepackage{caption}
\usepackage{xspace}
\usepackage{subcaption}
\usepackage{float}
\usepackage{qcircuit}
\definecolor{dkgreen}{rgb}{0,0.6,0}
\definecolor{gray}{rgb}{0.5,0.5,0.5}
\definecolor{mauve}{rgb}{0.58,0,0.82}
\begin{document}

\title{The simulation of distributed quantum algorithms}
\author{Sreraman Muralidharan$^1$}
\date{%
    $^1$Wells Fargo, New York, NY, USA\\%
}
\maketitle
\begin{abstract}
Distributed quantum computing (DQC) provides a way to scale quantum computers using multiple quantum processing units (QPU) connected through quantum communication links. In this paper, we have built a distributed quantum computing simulator and used the simulator to investigate quantum algorithms such as the quantum Fourier transform, quantum phase estimation, quantum amplitude estimation, and generation of probability distribution in DQC. The simulator can be used to easily generate and execute distributed quantum circuits, obtain and benchmark DQC parameters such as the fidelity of the algorithm and the number of entanglement generation steps, and use dynamic circuits in a distributed setting to improve results. We show the applicability of dynamic quantum circuits in DQC, where mid-circuit measurements, local operations, and classical communication are used in place of noisy inter-processor (non-local) quantum gates.
\end{abstract}

\section{Introduction}
\begin{figure}[h]
     \centering
     \begin{subfigure}[h]{0.3\textwidth}
         \centering
         \includegraphics[width=\textwidth]{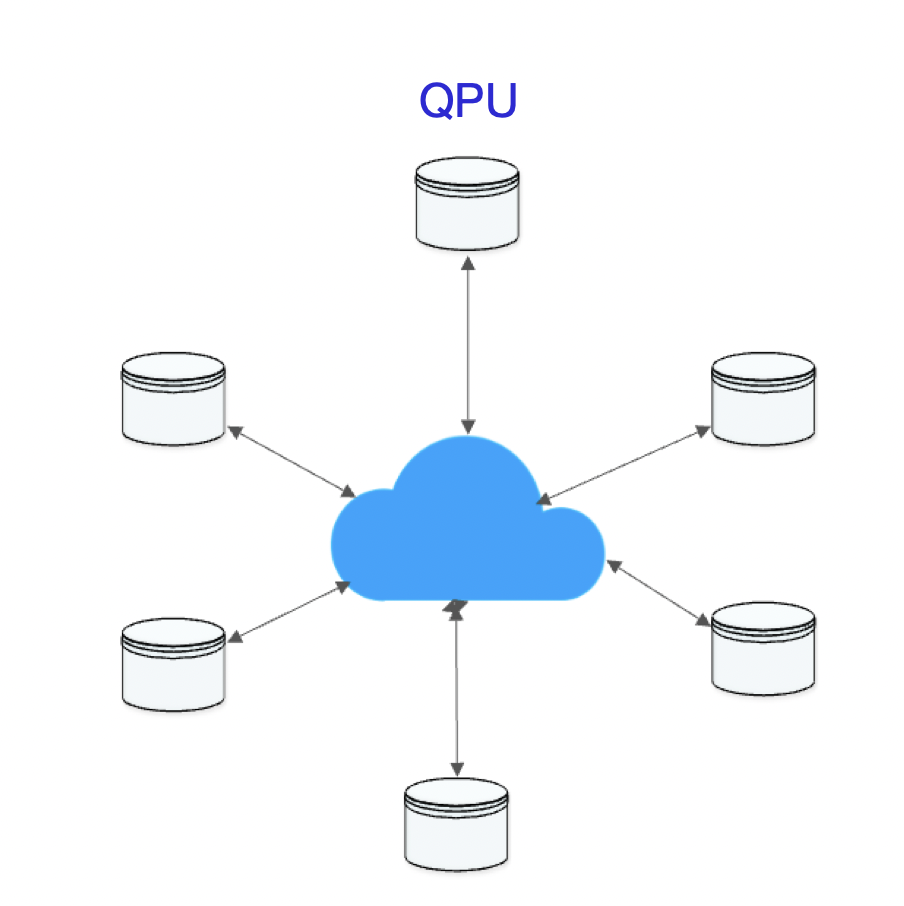}
         \caption{}
     \end{subfigure}
     \begin{subfigure}[h]{0.3\textwidth}
         \centering
         \includegraphics[width=\textwidth]{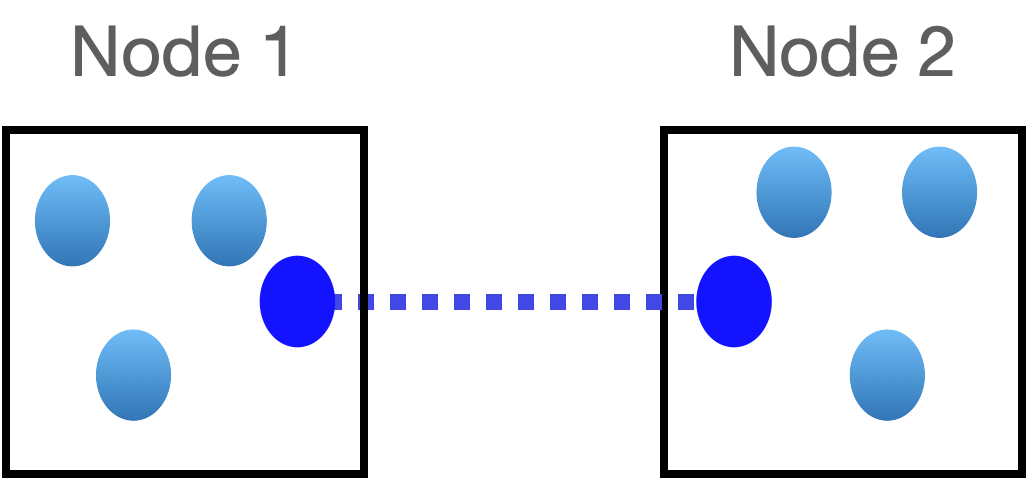}
         \caption{}
     \end{subfigure}
\caption{Schema for the DQC, a) Showing multiple quantum processing units. b) A two-node quantum computer showing entangled (dark blue) communication qubits and local (blue) processing qubits}
\label{fig:design}
\end{figure}

To make quantum computers advantageous in industry, the scalability of the
hardware needs to be significantly increased \cite{Buhrman2003, Beals2013, Ferrari2021, Parekh2021}. Due to the hardware limitations, increasing the qubits in the same device, increases the cross talk, decoherence that causes the effective gate and the measurement errors to increase \cite{Nicholas2025}.  Distributed quantum computing (DQC) \cite{Davis2023, Gyongyosi2021} offers an interesting approach, where instead of scaling up, we use multiple quantum processing units (QPUs) with fewer qubits connected through quantum communication links to increase computational capabilities \cite{Lim2005, Niu2023}. The QPUs can be separated by short distances, i.e., chiplets connected through microwave links \cite{Nicholas2025} or long distances connected through optical fibers/photonic links \cite{Main2025}. The nodes of the DQC (Fig. \ref{fig:design}) consist of the local qubits necessary for computation and communication qubits \cite{PhysRevLett.118.250502}, which are used to entangle the nodes and apply quantum gates between them \cite{Eisert2000, Monroe2009, Krutyanskiy2019, Main2025}.
The communication qubits can be entangled using a probabilistic entanglement generation procedure that relies on successful detector click patterns \cite{Nature2013, PhysRevLett.124.110501, PhysRevLett.120.200501}. The generation process is repeated until successful click patterns are obtained, indicating that the qubits are entangled. Entanglement generation between remote nodes has been experimentally realized in superconducting qubits  \cite{PhysRevLett.120.200501}, ions \cite{PhysRevLett.124.110501, Main2025}, and NV centers \cite{Nature2013} respectively. We will briefly study the probability of success of this procedure.
\par 
Suppose the channel's total efficiency (including qubit-node coupling) is $p$. For $k$ repetitions (trials), the probability of obtaining at least one entangled pair with one communication qubit/node is given by $P = (1-{(1-p)}^k)$. To increase the probability of success, we can have multiple communication ($m$) qubits per node, where the entanglement generation procedure is run for each pair in parallel. The probability that we have at least one entangled pair is given by $P_s = 1-(1-P)^m$.  
The entangled pair is used to apply a remote quantum gate between local qubits in the different nodes utilizing one ebit/CNOT gate by applying the quantum circuit shown in Fig. \ref{fig:non-local} \cite{Eisert2000}.
\begin{figure}[h]
\centering
\begin{center}
\scalebox{1.0}{
\Qcircuit @C=1.0em @R=0.2em @!R {
\lstick{q_0} & \qw & \qw & \ctrl{1} & \qw& \qw & \qw & \gate{Z({q_2})} & \qw\\
\lstick{q_1} & \qw & \qw & \targ & \qw & \qw & \meter & \cw \cwx[2]\\ 
\lstick{q_2} & \qw & \qw & \ctrl{1} & \gate{H}& \qw & \meter  & \cw \cwx[-2]\\
\lstick{q_3} & \qw & \qw & \targ & \qw & \qw& \qw & \gate{X({q_1})}& \qw\\
}}
\caption{The quantum circuit for the non-local teleportation-based CNOT gate between $q_0$ and $q_3$ using the Bell pair $q_1$ and $q_2$ prepared in $\frac{1}{\sqrt{2}}(|00\rangle+ |11\rangle$), followed by two local CNOT gates, measurement on the qubits $q_1$ and $q_2$, classical communication and conditional single-qubit quantum gates}
\label{fig:non-local}
\end{center}
\end{figure}
The non-local CNOT gate has been experimentally realized in superconducting qubits \cite{Chou2018} and ion qubits \cite{Daiss2021}. A non-local quantum circuit could be obtained using multiple non-local quantum gates, where the gates are applied sequentially through entangling, measuring, and re-entangling the same qubit with only one extra qubit/node.
Recently, distributed Grover's algorithm has been experimentally realized using dual species trapped ions connected through photonic link, where the nodes were separated by a distance of $2$ meters \cite{Main2025}. 
\par 
In this paper, we describe a distributed quantum computing simulator (DQCS) that uses non-local quantum gates between the nodes, creates distributed quantum circuits, and evaluates their performance in the presence of noise \cite{Eisert2000}. We use DQCS to investigate and benchmark the performance of several distributed quantum algorithms \cite{Zhou2023, Cirac1999, Yimsiriwattana2004}, demonstrating the applicability of dynamic quantum circuits in DQC that reduce resource overhead and achieve high fidelity. We illustrate this with a distributed quantum phase estimation (DQPE) example. We study the loading of the probability distribution in DQC, scaling with the number of nodes. We use DQCS to run the distributed quantum amplitude estimation (DQAE) algorithm on multiple nodes. The DQCS contains modules that provide users various ways to study, run, and evaluate the performance of quantum algorithms in DQC \cite {Cirac1999}. The DQCS could simulate and optimize the quantum network and circuit layers for distributed quantum algorithms. \par 
Several distributed quantum computing \cite{Ferrari2021, Rhea2021, Hanner2021, Cuomo2020} and quantum internet simulators \cite{Dahlberg_2019, Dahlberg_2022, QUISP} that have been developed that consider distributed quantum gates, compilers, and classical networks respectively. They consider the control platform for the hardware (Interlin-q) \cite{Rhea2021}, job schedulers \cite{Ferrari2021}, optimization using the quantum message parsing interface (QMPI) \cite{Hanner2021} for the simulation of DQC, analogous to the high performance computing. They consider the noiseless setting that does not include the gate, measurement imperfections, coupling efficiency, etc. that is crucial to realize DQC in the Noisy intermediate-scale quantum era (NISQ) quantum devices. In addition, they don't consider other techniques, such as dynamic quantum circuits in the distributed setting, that reduce the number of non-local quantum gates. The key challenge of the DQC is that each non-local two-qubit quantum gate requires two local two-qubit gates, two measurements, and efficient coupling between the nodes. Thus, for a given set of hardware parameters (such as gate fidelity, measurement error rate, and coupling efficiency), it's crucial to study the performance of a quantum algorithm \cite{PhysRevA.104.052404} in the DQC in the presence of various imperfections. The goal of the DQCS is to consider the implementation of DQC in the NISQ devices \cite{Main2025}, consider the hardware parameters including node efficiency (for entanglement generation), gate imperfections (for non-local quantum gates), and dynamic quantum circuits to improve fidelity, which are crucial for experiments. Further, since the DQCS is designed for the gate-model quantum computation, established techniques, such as circuit depth reduction \cite{Gyongyosi2020}, are crucial and can be applied to improve the time complexity of the distributed quantum algorithm. Additionally, the latency introduced by classical communication in the DQC can be minimized by optimizing qubit distribution \cite{Martinez2023}, utilizing routing protocols \cite{Hahn2019, Kristjánsson2024, Gyongyosi2022}, different protocols for entanglement generation \cite{Gyongyosi2022, Muralidharan2016} and entanglement swapping to maximize the rate, enhance the fidelity of the quantum algorithm \cite{Gyongyosi2018}.
\par
The paper is organized as follows: Section \ref{noise_model} introduces the noise model for non-local quantum gates. Section \ref{DQCS} introduces the DQCS using code snippets that enable us to obtain the distributed quantum circuit using the DQCS package. In sections \ref{dynamic}, \ref{dqpe}, we introduce dynamic quantum circuits for distributed quantum phase estimation and show that the advantage improves as we increase the number of qubits. In section \ref{dqae}, we estimate the resources for the distributed quantum amplitude estimation algorithm for uniform distribution and estimate the fidelity for the probability distribution loading of normal distribution for different numbers of nodes in the DQC.
\section{Noise model for non-local quantum gates}
\label{noise_model}
The noise model for non-local quantum gates can be studied using the depolarization channel $E$ that acts on the quantum state of $n$ qubits $\rho$, $\rho \rightarrow E(\rho)$ as
\begin{equation}
E(\rho) = (1-\lambda)\rho + \lambda Tr(\rho)\otimes\frac{I_{2^n}}{2^n},
\end{equation}
where $\lambda$ is the depolarization rate, $Tr(\rho)$ is the partial trace of $n$ qubits, and $I_{2^n}$ is the identity $2^n \times 2^n$ matrix, respectively. The single-qubit gate error channel $E$ can be obtained using the depolarization channel for $n=1$. Here, $\rho$ represents the quantum state before applying the single-qubit quantum gate ($U$) on the $i$th qubit. The channel gives $U_{i}\rho U_{i}^\dagger\rightarrow \rho'$,
\begin{equation}
\rho'= (1-\epsilon_d)U_{i}\rho U_{i}^\dagger + \epsilon_d Tr_{i}(\rho)\otimes\frac{I_2}{2},
\end{equation}
where $\epsilon_d$ is considered as the depolarization single-qubit gate error rate (respectively) and $Tr_{i}$ is the partial trace of the $i^{th}$ qubit. Consider a two-qubit quantum gate $U_{i,j}$ (between the qubits $i$ and $j$). The two-qubit quantum gate error channel could be obtained for $n=2$ that applies the transformation $U_{ij}\rho U_{ij}^\dagger\rightarrow \rho''$.
\begin{equation}
\rho'' = (1-\epsilon_g) U_{ij}\rho U_{ij}^\dagger + \epsilon_g Tr_{i,j}(\rho)\otimes\frac{I_4}{4}
\end{equation}
where $\epsilon_g$ is the two-qubit quantum gate error rate and $Tr_{i,j}$ is the partial trace of the qubits $(i,j)$. Note that in the case of a local CNOT gate, the fidelity from the error model is $\epsilon_g/2$. The fidelity of the quantum circuit after one successful implementation of the non-local CNOT gate (up to the first order) is $\epsilon \approx (\epsilon_d + \epsilon_g)$. If we consider $\epsilon_m$ to denote the measurement reset error, we may approximate that the fidelity for one distributed quantum gate is $(1- (\epsilon_d+\epsilon_g+2\epsilon_m))$. For $N$ distributed quantum gates, 
\begin{equation}
F \approx {(1- (\epsilon_d+\epsilon_g+2\epsilon_m))}^N
\end{equation}

\section{Distributed quantum computing simulator} \label{DQCS}
We built a distributed quantum computing simulator (DQCS) \cite{GitHub} that automates the creation of distributed quantum circuits for distributed quantum algorithms. It consists of the following modules:
\begin{itemize}
\item \textbf {Non-local quantum gates}: This module facilitates the application of non-local quantum gates between qubits residing in different nodes. It operates the quantum circuit in Fig. \ref {fig:non-local}, which enables the quantum circuit to span across nodes.
\item \textbf {Distributed circuits}: The distributed circuits module enables users to create distributed quantum circuits using a node dictionary and a quantum circuit. It automatically handles communication registers, so users don't need to specify them explicitly.
\item \textbf {Create distributed circuits}: To create the distributed quantum circuits for advanced quantum algorithms from existing quantum circuits, this module offers flexibility where the user can provide input to the communication register and obtain the distributed quantum circuit.
\item \textbf {Noise model}: We can experiment with the distributed quantum circuit in the presence of noise; the library offers an optional noise model module. This component incorporates realistic noise modeling, including single-qubit and two-qubit errors. The noise parameters can be changed to correspond to different experimental systems, and the fidelity of the quantum algorithm can be obtained. Note that the Bell pair is created using a CNOT gate in the simulator instead of the entanglement generation. Hence, there is a small difference in the noise model where the depolarization of the Bell pair is approximated with the gate error $\epsilon_g/2$. 
\end{itemize}

The steps to run the simulator are as follows:
\begin{enumerate}
\item Create a quantum circuit and instantiate the distributed circuits with your quantum circuit and the nodes to obtain the distributed quantum circuit.
\textit{}\begin{lstlisting}
import DQCS

qc = QuantumCircuit(3)
qc.h(0)
qc.h(1)
qc.cx(0, 2)
nodes = {"1": [0, 1], "2": [2]}
gate_app, circuit = DQCS.DistributedCircuits(qc, nodes).create_circuit()
\end{lstlisting}
The circuit gives the distributed quantum circuit the parameter $gate\textunderscore app = 1$ upon successfully creating the distributed quantum circuit.
\item Alternatively, use the $CreateDistributedCircuits$ instance with the quantum circuit, communication qubits, nodes, and the probability $p$ (coupling efficiency for the node) to create the distributed quantum circuit. Let us consider the example to illustrate the module.
\begin{lstlisting}
import DQCS

q_0 = QuantumRegister(2, 'a')
q_1 = QuantumRegister(2, 'b')
qc = QuantumCircuit(q_0, q_1)
nodes = {"1": [q_0[0], q_0[1]], "2": [q_1[0]], "3": [q_1[1]]}
qc.h(q_0[0])
qc.cx(q_0[1], q_1[1])

#Define an empty circuit with communication registers

comm = QuantumRegister(4, 'c')
c = ClassicalRegister(4, 'cl')
circ = QuantumCircuit(q_0, q_1, comm, c)
gate_app, circuit = DQCS.CreateDistributedCircuits(qc, c, circ, comm, nodes, 1).create_circuit()
\end{lstlisting}
The probability $p=1$ in the above example is the qubit-node coupling efficiency. This module is more suitable for running Qiskit's in-built quantum algorithms.
\item Instantiate the noise model with your distributed quantum circuit, specifying the number of shots, repetitions, and error probabilities (single, CNOT quantum gates, measurement, reset errors). We show an example of including a noisy $CNOT$ gate in the simulation.
\begin{lstlisting}
noise_model = noise.NoiseModel()
noise_model.add_all_qubit_quantum_error
(two_qubit_gate_error, ['cx'])
basis_gates = noise_model.basis_gates
hist = execute(qc, simulator,basis_gates,noise_model, shots).result()
\end{lstlisting}
\end{enumerate}
\section{Distributed quantum algorithms} \label{dqa}
Quantum algorithms are utilized to achieve speedup for various applications, including machine learning, optimization \cite{Farhi2014}, credit risk analysis \cite{Egger2021}, and option pricing using Monte Carlo integration \cite{Nik2020, Carrera2021, Herbert2022}. We consider the simulation of two distributed quantum algorithms, DQPE and DQAE, in  DQC, which are the underlying quantum algorithms for more advanced quantum algorithms \cite{Egger2021, Nik2020, Carrera2021, Herbert2022}.
\subsection{Dynamic quantum circuits} \label{dynamic}
Dynamic quantum circuits involve using mid-circuit measurements, qubit reset, reuse, and single-qubit conditional quantum gates \cite{PhysRevLett.127.100501}. Here, we explore using dynamic quantum circuits instead of distributed quantum gates in the DQC  \cite{dynamic}. One of the examples of dynamic quantum circuits is when we have a quantum Fourier transform followed by measurement \cite{dynamic}.
\par 
Let's consider a quantum circuit for the QFT on three qubits in Fig. \ref{fig:qft_1}, followed by measurement of the qubits. 
\begin{figure}[H]
\centering
\small
\begin{center}
\scalebox{1.0}{
\Qcircuit @C=1.0em @R=0.8em @!R {
\lstick{q_0} & \gate{H} & \ctrl{1} & \dstick{\hspace{2.0em}P(-\pi/2)} \qw & \qw & \qw & \qw & \ctrl{2} & \qw & \qw & \qw & \qw & \qw & \qw & \qw & \qw & \qw & \meter\\
\lstick{q_1} & \qw & \control \qw & \qw & \qw & \qw & \gate{H} & \qw & \dstick{\hspace{2.0em}P(-\pi/4)} \qw & \qw & \qw & \ctrl{1} & \dstick{\hspace{2.0em}P(-\pi/2)} \qw & \qw & \qw & \qw& \qw& \meter\\
\lstick{q_2} & \qw & \qw & \qw & \qw & \qw & \qw &\control \qw &  \qw & \qw &  \qw &\control \qw &   \qw & \qw & \qw& \gate{H} \qw & \qw & \meter
}}
\caption{Quantum Fourier transform on three qubits, where $P(\theta)$ refers to the phase gate}
\label{fig:qft_1}
\end{center}
\end{figure}
Fig. \ref{fig:dqft_1} shows the corresponding dynamic quantum circuit.
\begin{figure}[H]
\centering
\begin{center}
\scalebox{1.0}{
\Qcircuit @C=1.0em @R=0.8em @!R {
\lstick{q_0} & \gate{H} & \meter\\
\lstick{q_1} & \qw & \gate{P(-\pi/2)} \cwx[-1] & \gate{H} & \meter\\
\lstick{q_2} & \qw & \gate{P(-\pi/4)} \cwx[-1] & \qw & \gate{P(-\pi/2)} \cwx[-1] & \gate{H} & \meter
}}
\caption{Dynamic quantum Fourier transform on three qubits}
\label{fig:dqft_1}
\end{center}
\end{figure}
Note that the dynamic QFT does not suffer from two-qubit gate errors; it suffers from depolarization errors $\epsilon_d = 1-e^{-t/T_2}$ \cite{Piparo2024}, where t is the classical communication time, and $T_2$ is the coherence time of the qubit. For $T_2=50\mu s$, QPUs separated by $10m$, speed of light in fiber $c = 2\times 10^{8}m/s$, we get $\epsilon_d \approx 10^{-3}$. We can now illustrate the application of distributed dynamic QFT to the quantum phase estimation. 
\subsection{Distributed dynamic quantum phase estimation} \label{dqpe}
 The problem of quantum phase estimation (QPE) is to determine the eigenvalue of the unitary gate that acts as $U|\psi\rangle = e^{2i\pi y}|\psi\rangle$, where $|\psi\rangle$ is the eigenstate of $U$, and $y$ is a fraction corresponding to the eigenvalue. The quantum circuit used to implement QPE is shown in Fig. \ref{fig:dqpe_1}.
\begin{figure}[h]
\centering
\begin{center}
\scalebox{0.8}{
\Qcircuit @C=0.6em @R=0.8em @!R {
\lstick{|0\rangle} & \gate{H} & \qw & \qw & \ctrl{3} & \gate{H} & \meter\\
\lstick{|0\rangle} & \gate{H} & \qw & \ctrl{2} & \qw & \qw & \gate{P(-\pi/2)} \cwx[-1] & \gate{H} & \meter\\
\lstick{|0\rangle} & \gate{H} & \ctrl{1} & \qw & \qw & \qw & \gate{P(-\pi/4)} \cwx[-1] & \qw & \gate{P(-\pi/2)} \cwx[-1] & \gate{H} & \meter\\
\lstick{|\psi\rangle} & \qw & \gate{U^{2^2}} & \gate{U^{2^1}} & \gate{U^{2^0}} & \qw & \qw & \qw & \qw & \qw & \qw
}}
\caption{Distributed dynamic quantum phase estimation, where the qubits are in different nodes. The controlled unitary gates are applied using DQCS, followed by dynamic QFT}
\label{fig:dqpe_1}
\end{center}
\end{figure}
We can derive the fraction $y$ from the outcome of the measurement. For instance, if we obtain the measurement outcome $y_0y_1y_2$, where each $y_i \in {0,1}$, then the fraction $y$ is calculated as $(y_0y_1y_2)_{10}/2^3$. To illustrate, let's examine the implementation of QPE for the unitary gate $U = R_X(\pi) \otimes R_Y(\pi)$ acting on two qubits. One of the eigenvalues for this operator is $-1$ ($y = 1/2$), corresponding to a measurement output of $"100"$. We prepare the eigenstate $|\psi\rangle$, create the distributed QPE (DQPE) for the following qubit distributions: 1: [0, 1, 2], 2: [3, 4], and apply the dynamic QFT to the distributed quantum circuit. The probability of the measurement outcome $"100"$ is shown in Fig. \ref{fig:qpe_1}, where we consider the locally controlled unitary gates and quantum Fourier transform for the QPE (red) (Fig. \ref{fig:qft_1}). For the DQPE (blue), we consider distributed two-qubit quantum gates and dynamic QFT without two-qubit quantum gates for the quantum Fourier transform (Fig. \ref {fig:dqft_1}). 
The probability of measuring $"100"$ is shown in Fig. \ref{fig:qpe_1}. For the DQPE (blue), we consider distributed two-qubit quantum gates and dynamic QFT without two-qubit quantum gates for the quantum Fourier transform (Fig. \ref {fig:dqft_1}). We consider the gate error $\epsilon_g$, depolarization error $\epsilon_d$, and the measurement error $\epsilon_m$. In Fig. \ref{fig:qpe_1}, we consider the following scenarios
\begin{enumerate}
    \item  Probability vs $\epsilon_g$, where $\epsilon_g = 10\epsilon_d$, $\epsilon_m=0$ (Fig. \ref{fig:1a}).
    \item Probability vs $\epsilon_g$, where $\epsilon_g = 10\epsilon_d$, $\epsilon_m=\epsilon_g/4$ (Fig. \ref{fig:1b}).
\end{enumerate}
While each distributed quantum gate requires two local CNOT gates, the dynamic quantum Fourier transform doesn't suffer from two-qubit quantum gate errors, only measurement and depolarization errors, giving better results in Fig. \ref{fig:qpe_1}. Since each non-local quantum gate requires measurement that is not present in the local implementation of a quantum circuit, we consider the measurement error in our analysis. Note that there are efficient methods to reduce the measurement error to $\epsilon_g/4$ \cite{Knill2005, Muralidharan2016}.
\begin{figure}[H]
     \begin{subfigure}[b]{0.5\textwidth}
         \centering
         \includegraphics[width=\linewidth]{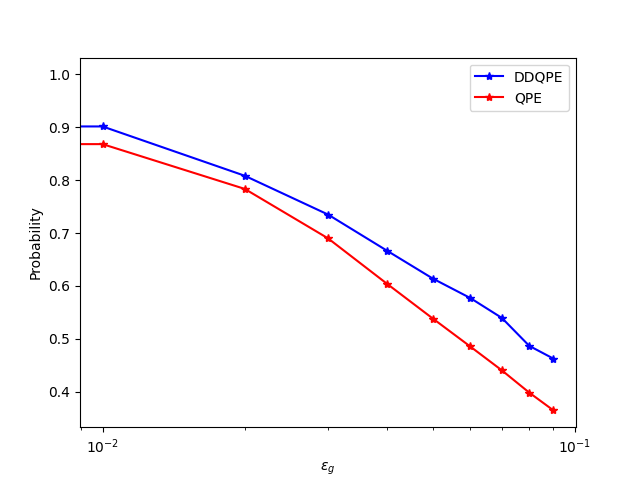}
         \caption{Probability vs $\epsilon_g$,  $\epsilon_d = \epsilon_g/10$, $\epsilon_m=0$}
         \label{fig:1a}
     \end{subfigure} 
    \hfill
    \begin{subfigure}[b]{0.5\textwidth}
         \centering
         \includegraphics[width=\linewidth]{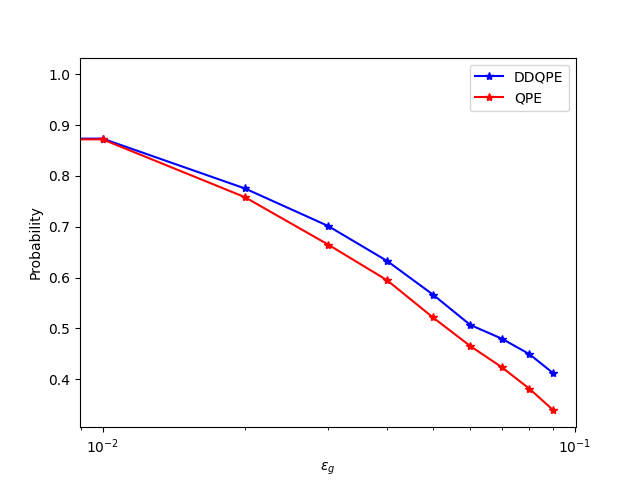}
         \caption{Probability vs $\epsilon_g$,  $\epsilon_d = \epsilon_g/10$, $\epsilon_m=\epsilon_g/4$}
         \label{fig:1b}
     \end{subfigure}
        \caption{The probability measuring the required bit for the distributed QPE (DQPE) (blue) vs. QPE (red) in the presence of depolarization, gate, measurement errors $\epsilon_d$, $\epsilon_g$, $\epsilon_m$ respectively.}
        \label{fig:qpe_1}
\end{figure}
The improvement obtained with DQPE increases when considering a larger number of qubits.
Let us consider the inverse QFT on five qubits (Fig. \ref{fig:qft}) that uses ten two-qubit and five Hadamard gates. 
\begin{figure}[H]
\begin{center}
\small
\scalebox{0.6}{
\Qcircuit @C=0.5em @R=0.5em {
\lstick{q_0} & \gate{H} & \ctrl{1} & \qw & \ctrl{2} & \qw & \qw & \ctrl{3} & \qw & \qw & \qw & \ctrl{4} & \qw & \qw & \qw & \qw &\meter \\
\lstick{q_1} & \qw &  \gate{P(-\pi/2)} & \gate{H} & \qw & \ctrl{1} & \qw & \qw & \ctrl{2} & \qw & \qw & \qw & \ctrl{3} & \qw & \qw &\qw & \meter \\
\lstick{q_2} & \qw & \qw & \qw & \gate{P(-\pi/4)} & \gate{P(-\pi/2)} & \gate{H} & \qw & \qw & \ctrl{1} & \qw & \qw & \qw & \ctrl{2} & \qw & \qw &\meter \\
\lstick{q_3} & \qw & \qw & \qw & \qw & \qw & \qw & \gate{P(-\pi/8)} & \gate{P(-\pi/4)} & \gate{P(-\pi/2)} & \gate{H} & \qw & \qw & \qw & \ctrl{1} & \qw & \meter \\
\lstick{q_4} & \qw & \qw & \qw & \qw & \qw & \qw & \qw & \qw & \qw & \qw & \gate{P(-\pi/16)} & \gate{P(-\pi/8)} & \gate{P(-\pi/4)} & \gate{P(-\pi/2)} & \gate{H} & \meter
}}
\caption{Quantum Fourier transform (inverse) on five qubits, where $P(\theta)$ refers to the phase gate}
\label{fig:qft}
\end{center}
\end{figure}
The dynamic quantum circuit is given in Fig. \ref{fig:dqft},
\begin{figure}[H]
\centering
\begin{center}
\scalebox{0.9}{
\Qcircuit @C=1.0em @R=0.8em @!R {
\lstick{q_0} & \gate{H} & \meter\\
\lstick{q_1} & \qw & \gate{P(-\pi/2)} \cwx[-1] & \gate{H} & \meter\\
\lstick{q_2} & \qw & \gate{P(-\pi/4)} \cwx[-1] & \qw & \gate{P(-\pi/2)} \cwx[-1] & \gate{H} & \meter \\
\lstick{q_3} & \qw & \gate{P(-\pi/8)} \cwx[-1] & \qw & \gate{P(-\pi/4)} \cwx[-1] & \qw & \gate{P(-\pi/2)} \cwx[-1] & \gate{H} & \meter \\
\lstick{q_4} & \qw & \gate{P(-\pi/16)} \cwx[-1] & \qw & \gate{P(-\pi/8)} \cwx[-1] & \qw & \gate{P(-\pi/4)} \cwx[-1] & \qw& \gate{P(-\pi/2)} \cwx[-1] & \gate{H} &\meter\\
}}
\caption{Dynamic quantum Fourier transform using five qubits}
\label{fig:dqft}
\end{center}
\end{figure}
The quantum circuit used to implement the DQPE is shown in Fig. \ref{fig:dqpe}.
\begin{figure}[h]
\centering
\begin{center}
\scalebox{0.6}{
\Qcircuit @C=1.0em @R=0.8em @!R {
  \lstick{|0\rangle} & \gate{H} & \qw & \ctrl{5} & \qw   & \qw & \gate{H} &   \qw & \meter \\
  \lstick{|0\rangle} & \gate{H} & \qw & \qw & \ctrl{4} & \qw & \qw &  \qw &\gate{P(-\pi/2)} \cwx[-1] & \gate{H} & \meter \\
  \lstick{|0\rangle} & \gate{H} & \qw & \qw & \qw & \ctrl{3} & \qw &  \qw &\gate{P(-\pi/4)} \cwx[-1] & \qw & \gate{P(-\pi/2)} \cwx[-1] & \gate{H} & \meter \\
  \lstick{|0\rangle} & \gate{H} & \qw & \qw & \qw & \qw & \ctrl{2} &  \qw &\gate{P(-\pi/8)} \cwx[-1] & \qw & \gate{P(-\pi/4)} \cwx[-1] & \qw & \gate{P(-\pi/2)} \cwx[-1] & \gate{H} & \meter \\
  \lstick{|0\rangle} & \gate{H} & \qw & \qw & \qw & \qw & \qw & \ctrl{1} &  \gate{P(-\pi/16)} \cwx[-1] & \qw & \gate{P(-\pi/8)} \cwx[-1] & \qw & \gate{P(-\pi/4)} \cwx[-1] & \qw & \gate{P(-\pi/2)} \cwx[-1] & \gate{H} & \meter \\
  \lstick{|\psi\rangle} & \qw & \qw & \gate{U^{2^4}} & \gate{U^{2^3}} & \gate{U^{2^2}} & \gate{U^{2^1}} & \gate{U^{2^0}} & \qw & \qw & \qw & \qw
}}
\caption{Distributed dynamic quantum phase estimation for five qubits where the qubits are in different nodes. The controlled unitary gates are applied using DQCS, followed by dynamic QFT}
\label{fig:dqpe}
\end{center}
\end{figure}\\
We estimate the fraction $y$ that is calculated as $(y_0y_1y_2y_3y_4)_{10}/2^5$. We prepare the eigenstate $|\psi\rangle$, create the distributed QPE (DQPE) for the following qubit distributions: 1: [0, 1, 2, 3, 4], 2: [5, 6], and apply the dynamic QFT to the distributed quantum circuit. We consider the following scenarios:
\begin{enumerate}
    \item Probability of obtaining the correct outcome vs error rate $\epsilon$, where $\epsilon_g = \epsilon_d = \epsilon$, $\epsilon_m=0$ (Fig. \ref{fig1a}).
    \item  Probability vs the two-qubit gate error $\epsilon_g$, where $\epsilon_g = 10 \epsilon_d$, $\epsilon_m=0$ (Fig. \ref{fig1b}).
    \item Probability vs $\epsilon$, where $\epsilon_g = \epsilon_d = \epsilon$, $\epsilon_m=\epsilon_g/4$ (Fig. \ref{fig1c}).
    \item Probability vs $\epsilon_g$, where $\epsilon_g = 10 \epsilon_d$, $\epsilon_m=\epsilon_g/4$ (Fig. \ref{fig1d}).
\end{enumerate}
In Fig. \ref{fig:qpe}, we note that there is better improvement obtained using DQPE (using five qubits) than three qubits (Fig. \ref{fig:qpe_1}). This could be understood theoretically as follows: For quantum phase estimation using $n$ control qubits, suppose that we need $N$ two-qubit quantum gates to realize the controlled-unitary operations, and the quantum Fourier transform requires $n(n-1)/2$ two-qubit quantum gates. Theoretically, assuming the same error for the two-qubit (controlled-unitary and controlled phase) gates, the fidelity of the algorithm is approximated $F \approx (1-\epsilon_g)^{N+\frac{n}{2}(n-1)}$. For DQPE, where we have $N$ distributed quantum gates (each applied using two CNOT quantum gates), the fidelity is given by $F \approx (1-\epsilon_g)^{2N}$.
\begin{figure}[H]
     \begin{subfigure}[b]{0.5\textwidth}
         \centering
         \includegraphics[width=\linewidth]{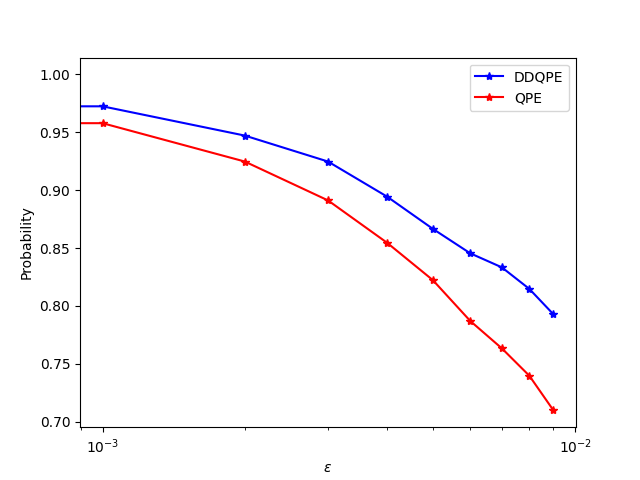}
         \caption{$\epsilon_g = \epsilon_d = \epsilon$, $\epsilon_m=0$}
         \label{fig1a}
     \end{subfigure}
     \hfill
     \begin{subfigure}[b]{0.5\textwidth}
         \centering
         \includegraphics[width=\linewidth]{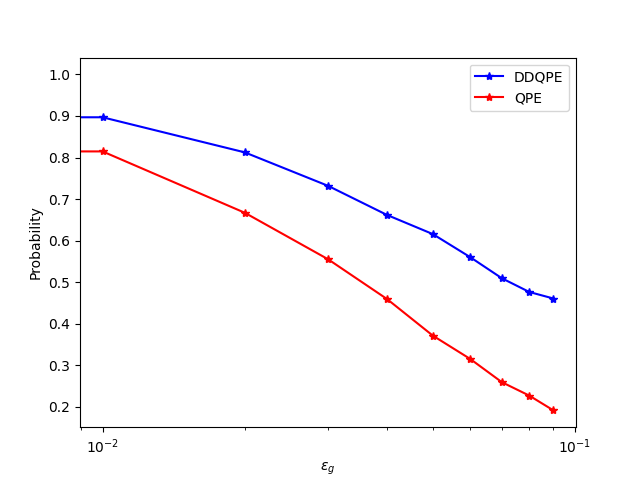}
         \caption{$\epsilon_d=\epsilon_g/10$, $\epsilon_m=0$}
         \label{fig1b}
     \end{subfigure} \\
     \hfill
     \begin{subfigure}[b]{0.5\textwidth}
         \centering
         \includegraphics[width=\linewidth]{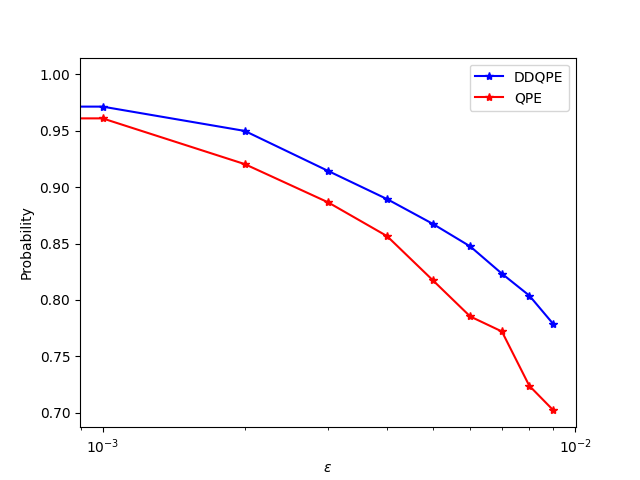}
         \caption{$\epsilon_g = \epsilon_d= \epsilon$, $\epsilon_m=\epsilon_g/4$}
         \label{fig1c}
     \end{subfigure}
    \begin{subfigure}[b]{0.5\textwidth}
         \centering
         \includegraphics[width=\linewidth]{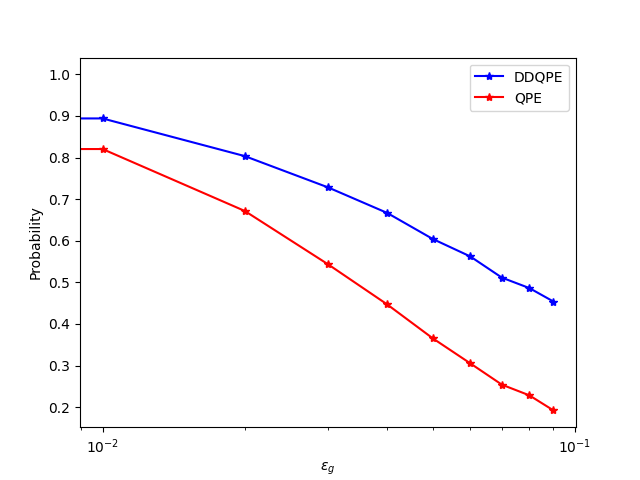}
         \caption{$\epsilon_d=\epsilon_g/10$, $\epsilon_m=\epsilon_g/4$}
         \label{fig1d}
     \end{subfigure}
        \caption{The probability of the correct outcome for the distributed QPE (DQPE) (blue) vs. QPE (red) in the presence of gate, depolarization, and measurement errors}
        \label{fig:qpe}
\end{figure}
\subsection{Distributed quantum amplitude estimation} \label{dqae}
Quantum amplitude estimation (QAE) is used to calculate the unknown amplitude of a quantum state. For a given unitary operator $A$,
\begin{equation}
A|0\rangle_{n+1} = \sqrt{1-a} |\psi_0\rangle_n |0\rangle + \sqrt{a} |\psi_1\rangle_n |1\rangle,
\end{equation}
where $a$ is the unknown amplitude to be estimated, and $|\psi_0\rangle_n$ and $|\psi_1\rangle_n$ are (normalized) good and bad states, respectively, that correspond to the probability of measurement of $|0\rangle$ and $|1\rangle$ in the last qubit, respectively. 
\par 
QAE has applications such as calculating the expectation value of the discrete linear function, quantum Monte Carlo integration \cite{Nik2020, Herbert2022}, quantum risk analysis, and portfolio optimization. Using multiple nodes, we use the DQCS to simulate distributed QAE (DQAE). There are various approaches to performing QAE. Here, we study DQAE using maximum likelihood estimation \cite{Suzuki2020}, which achieves quadratic speedup over classical Monte Carlo algorithms in DQC without using the QFT. We show that the estimation error for the DQAE \cite{Suzuki2020}, simulated using the DQCS, achieves the Cramer-Rao bound, which provides a way to scale QAE using multiple QPUs. 

\begin{figure}[H]
\centering
\includegraphics[width=8cm]{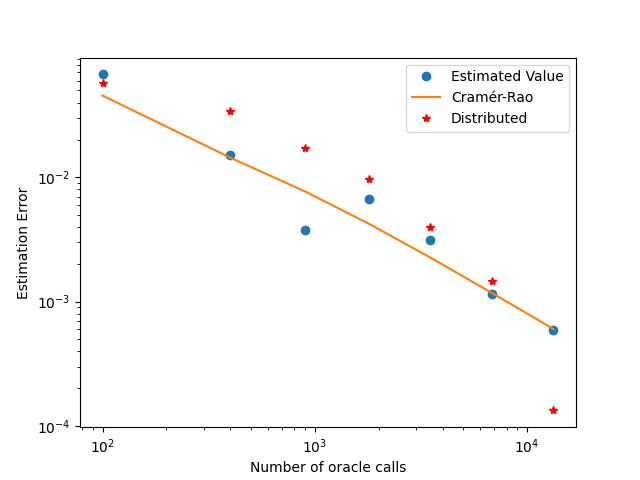}
\caption{The estimated value in a quantum computer (1 node), distributed quantum computer (2 nodes), and Cramer-Rao bound \cite{Suzuki2020}}
\label{fig:cramer-rao}
\end{figure}

We prepare the $A$ operator, which loads a discrete probability distribution to a quantum state using the probability distribution loading $P(x)$ \cite{Nik2020}. The bounded function $f(x)$ is constructed from the samples of $P(x)$ and realized using multiple controlled-$R_y$ gates with controlled-$R_y(f(i))$ on the gate on the $ith$ qubit \cite{Nik2020}. The probability of the measurement of $|1\rangle$ on the last qubit gives the expectation value $\mathrm{E}[f(x)]$.
\begin{equation}
\mathrm{E}[f(x)] = \sum_{i=0}^{2^{n}-1} p(x_i) f(x_i),
\end{equation}
Suppose P(x) is a uniform distribution, the probability of measuring $|1\rangle$ on the last qubit is given by $\frac{1}{2^n}\sum_{i=0}^{2^{n}-1}f(x_i)$ for $x_i=i/2^n$ \cite{Carrera2021}. For QAE, we construct the $Q$ operator, such that $Q = -AS_0A^{-1}S_X$. The operator $S_X$ leaves the good states unchanged and includes a phase $(-1)$ on the bad states. $S_X|\psi_0\rangle|0\rangle = |\psi_0\rangle|0\rangle$, $S_X|\psi_1\rangle|1\rangle = -|\psi_1\rangle|1\rangle$. The operator $S_0$ attaches a phase (-1) to $|0\rangle_n$ and leaves other basis states unchanged; the repeated application of $Q$, $M$ times, on the state $A|0\rangle$ gives
\begin{equation}
Q^M |\psi\rangle = \cos(2M+1)\theta |\psi_0\rangle |0\rangle + \sin(2M+1)\theta |\psi_1\rangle |1\rangle,
\end{equation}
QAE requires multiple controlled unitary gates to implement the reflection operators. We could write multiple CNOT gates (with $n$ control qubits and a single-qubit unitary) with single-qubit gates and $\leq (20n-38)$ CNOT gates \cite{Rv2023, Eisert2000}. We consider the QAE without phase estimation that uses the maximum likelihood estimate \cite{Suzuki2020} to integrate $\int_0^c \sin^2(x)$ in DQC. We compare the estimated error values against the Cramer-Rao bound \cite{Suzuki2020}, where $P(x)$ is a uniform distribution obtained by applying Hadamard gates to all qubits. The function $f(x)$ is prepared with $a = c/2^{n-1}, b = c/2^n$ using 1 $R_y$ gate and $n$ controlled-Ry gates, $Ry(c/2^{n-i})$ \cite{Suzuki2020}. It can be seen in Fig. \ref {fig:cramer-rao} (($c=\frac{\pi}{3}$)) that DQAE can be simulated and the estimated value achieves the Cramer-Rao bound.
\par
\begin{figure}[h]
\centering
\includegraphics[width=8cm]{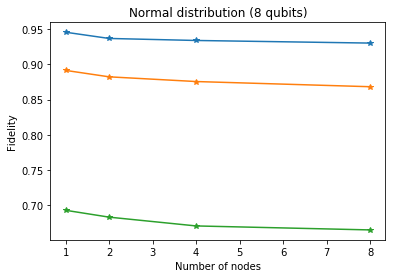}
\caption{The Heilinger fidelity of a normal distribution with eight qubits with different numbers of nodes, where $\epsilon_d = \epsilon_g = (0.002 (\text{blue}), 0.005 (\text{orange}), 0.009 (\text{green}))$}
\label{fig:fidelity}
\end{figure}

 
In our simulation, we used Qiskit's transpile function to apply a distributed multiple controlled gate and distributed unitary gates using non-local CNOT gates (Fig. \ref{fig:non-local}). The DQCS can be used to obtain the resource estimation for the DQAE for various node parameters in Table \ref{tab:table_1}.
\begin{table*}[t]
    \centering
    \begin{tabular}{|c|c|c|c|}
    \hline
     $\text{n}_{\text{nodes}}$ & $\text{n}_{\text{Grover}}$ & $k$ & $\text{Estimation \ error}$ \\
     \hline
     (2,3) & $0$ & [(6,7,9),(6,7,11)] & (0.00881913, 0.01116442) \\ \hline
     (2,3) & $1$ & [(28,30,48),(30,34,54)] & (0.00936860, 0.02658211) \\ \hline
     (2,3) & $2$ & [(50,59,91),(54,64,101)] & (0.00576238, 0.00911406) \\ \hline
     (2,3) & $4$ & [(94,119,157),(102,113,193)] & (0.00498004, 0.00153080) \\ \hline
     (2,3) & $8$ & [(182,220,323),(198,238,338)] & (0.00251507, 0.00070526) \\ \hline
     (2,3) & $16$ & [(358,441,687),(390,449,758)] & (0.00112920, 0.00141048) \\ \hline
     (2,3) & $32$ & [(710,900,1366),(774,908,1371)] & (0.00050303, 0.00017524) \\ \hline
\end{tabular}
    \caption{Resource estimation for the distributed quantum amplitude estimation (DQAE) obtained using the DQCS where the number of nodes $\text{n}_{\text{nodes}}$, number of Grover operators $\text{n}_{\text{Grover}}$,  number of entanglement generation steps $k$ for the qubit-node coupling parameter $p=(1, 0.8, 0.5)$, and estimation error are considered}
    \label{tab:table_1}
\end{table*}
We could also consider QAE using normal distribution instead of uniform distribution. It is crucial to understand the quantum state preparation of probability distribution in DQC. The normal distribution
\begin{equation}
P(X=x) = \frac{1}{\sqrt{2\pi\sigma^2}} e^{\frac{-(x-\mu)^2}{\sigma^2}},
\end{equation}
loaded to quantum state $|p\rangle = \sum_{i=0}^{2^{d-1}} \sqrt{p_i}|x_i\rangle$, where $p_i$ is sampled from $P(x)$ with $2^d$ points \cite{Nik2020, Carrera2021}. We use the DQCS to simulate the quantum state in preparation for a probability distribution. We consider a normal distribution with $8$ qubits with the mean $\mu=1$ and $\sigma=2$ in the DQCS, with $n=1,2,4,8$ nodes, to obtain the Heilinger fidelity in Fig. \ref{fig:fidelity}.
\section{Conclusion}
We outlined a distributed quantum circuit simulator (DQCS), which uses multiple interconnected quantum processing units (QPUs) that consist of local and communication qubits. We studied foundational quantum algorithms such as the Fourier transform, quantum phase estimation, and amplitude estimation using DQCS and showed the application of dynamic quantum circuits in DQC. Recently, there has been growing interest in the realization of quantum networks and distributed quantum computing in various platforms such as superconducting qubits, NV centers \cite{Nature2013, NV}, ions \cite{Main2025, PhysRevLett.124.110501}, and neutral atoms \cite{neutral}. The DQCS provides a platform for users to input the experiment parameters, simulate distributed quantum circuits, and obtain the estimates for quantum algorithms, including quantum phase estimation, quantum amplitude estimation, quantum state preparation of the probability distribution, etc. In the future, we aim to integrate quantum internet simulators with the DQCS to study distributed quantum algorithms in the quantum internet. We look forward to include quantum error correction and fault tolerance in the DQCS to obtain favorable polynomial scaling for the fidelity of quantum algorithms \cite{Muralidharan2016}. We could also study distributed quantum machine learning algorithms (QML) for classification and regression \cite{Tang2023}, federated QML \cite{Sam2021} for quantum neural networks (QNN) \cite{Gyongyosi2019}, dynamic QML for recurrent QNN's \cite{Gyongyosi2019}.
\section*{Acknowledgements}
I want to thank Constantin Gonciulea, Charlee Stefanski, Vanio Markov, and David Novak for their helpful discussions, comments on the DQCS, and suggestions for improving the quality of the manuscript.
\par The views expressed in this article are those of the authors and do not represent the views of Wells Fargo. This article is for informational purposes only. Nothing contained in this article should be construed as investment advice. Wells Fargo makes no express or implied warranties and expressly disclaims all legal, tax, and accounting implications related to this article.
\printbibliography 
\end{document}